\documentclass[twocolumn,pra,showpacs,superscriptaddress]{revtex4}
\usepackage{amssymb}
\usepackage{amsmath}
\usepackage{graphicx}
\usepackage{subfigure}
\usepackage{epsfig}
\usepackage{amsfonts}
\usepackage{CJK}
\usepackage{subfigure}
\usepackage{xcolor}
\usepackage{natbib}

\definecolor{red}{rgb}{1,0,0}
\definecolor{blue}{rgb}{0,0,1}

\begin{document}

\title{Dynamics of Quantum Zeno and Anti-Zeno Effects in  Open System}

\author{Peng Zhang }
\affiliation{Department of Physics, Renmin University of China,
Beijing, 100190, China}
\author{Qing Ai }
\affiliation{Institute of Theoretical Physics, Chinese Academy of
Sciences, Beijing, 100190, China}
\author{Yong Li }
\affiliation{Beijing Computational Science Research Center, Beijing, 100084, China}
\author{D. Z. Xu }
\affiliation{Institute of Theoretical Physics, Chinese Academy of
Sciences, Beijing, 100190, China}%
\author{C. P. Sun }
\email{suncp@itp.ac.cn}%
\affiliation{Institute of Theoretical Physics, Chinese Academy of
Sciences, Beijing, 100190, China}%

\begin{abstract}
We provide a general dynamical approach for the quantum Zeno
and anti-Zeno effects in an open quantum system under repeated
non-demolition measurements.
In our approach the repeated measurements are described by a general
dynamical model without
the wave function collapse postulation. Based on that model, we further study both the
short-time and long-time evolutions of the open quantum system under
repeated non-demolition measurements, and derive the measurement-modified decay rates of the
excited state. In the cases with frequent ideal measurements at zero-temperature, we
re-obtain the same decay rate as that from the wave function
collapse postulation (Nature {\bf 405}, 546 (2000)). The correction
to the ideal decay rate is also obtained under the non-ideal
measurements. Especially, we find that  the quantum Zeno and
anti-Zeno effects are possibly {\it enhanced} by the non-ideal
natures of  measurements. For the open system under measurements with
arbitrary period, we generally
derive the rate equation for the long-time evolution for the cases with {\it arbitrary} temperature and noise spectrum,
and show that in the long-time evolution the noise spectrum
is effectively tuned by the repeated measurements.
Our approach is also able to describe the quantum Zeno
and anti-Zeno effects given by the phase modulation pulses, as well
as the relevant quantum control schemes.
\end{abstract}

\maketitle

\section{Introduction}

Quantum Zeno effect (QZE) \cite{QZ} and quantum anti-Zeno effect (QAZE) \cite%
{QAZ} are among the most interesting results given by quantum mechanics. The
two effects describe the evolution of a quantum system under repeated (or
continuous) measurements. QZE shows that, for a closed quantum system with
finite-dimensional Hilbert space, the unitary evolution can be inhibited by
the repeated measurements. Not surprisingly, the similar inhibition can also
occurs in the Rabi oscillation of a quantum open system coupled with an
environment.

On the other hand, for an unstable state in an open quantum system with
dissipation, QZE shows that the time-irreversible decay of such a state can
be suppressed when the quantum measurements are {frequent enough}.
In addition to the QZE in such {an} 
extreme limit, the decay rate of the unstable state in a dissipative system
is also possible to be increased if the measurements are repeated with an
\textquotedblleft intermediate" frequency. That is known as quantum
anti-Zeno effect. QZE and QAZE have attracted much attention since they were
proposed. So far the two effects have been experimentally observed in the
systems of trapped ions \cite{ExpIonTrap1,ExpIonTrap2}, ultracold atoms \cite%
{ExpColdAtom1,ExpColdAtom2,ExpColdAtom3}, molecules \cite{ExpMolecule} and
cavity quantum electric dynamics (cavity QED) \cite{ExpCQED}.

The QZE and QAZE were previously derived with the postulation of wave packet
collapse in the quantum measurement. After the initial proposals of the two
effects, many authors have discussed the dynamical {explanation} of QZE and
QAZE via the unitary quantum mechanical approach without wave packet
collapse. For the unitary evolution of a closed system or the Rabi
oscillation of an open system, many authors pointed out that the QZE in
these systems can also be explained in a dynamical approach \cite%
{DQZE1990,DQZE1990b,DQZE1990c,
DQZE1991,DQZE1991b,DQZE1991c,DQZE1991d,DQZE1992,DQZE1993,DQZE1993b,DQZE1993c, DQZE1994,DQZE1994b,
DQZE1994c,DQZE1994d,DQZE1994e,DQZE1995, DQZE1996,DQZE1997, DQZE1998,DQZE2010,DQAZE2005}%
. These explanations are based on the considerations of the quantum dynamics
of the quantum system as well as the apparatuses during the measurements. In
these complete dynamical analysis, QZE can appear naturally during the evolution of the reduced
density matrix of the system under the frequent interactions with the
apparatuses, and the postulation of wave packet collapse is not required.
These explanations of QZE were previously given for many
special systems and set-ups for measurements (e.g., two-level atoms under
the detection with laser pulses). Recently some of us (D. Z. Xu, Qing. Ai and C. P. Sun)
provided a general dynamical proof of QZE for any closed system under
quantum non-demolition measurements \cite{DQZE2010}.

For the QZE and QAZE {of the decay of} 
an unstable state in a quantum open system with dissipation, there are also
similar discussions in which the dynamical process of the measurements is
included \cite{DQAZEhalf2002,DQAZEhalf2006,DQAZE2005,
DQAZEhalf2001,DQAZEhalf2001b,DQAZEhalf2004,DQAZE2010}. In Refs. \cite%
{DQAZEhalf2001,DQAZEhalf2001b,DQAZEhalf2004,DQAZEhalf2006}, where the
quantum open system is also assumed to be under repeated measurements, the
dynamics of every individual measurement is discussed in detail.
Nevertheless, in these references the total survival probability of the
initial state of the open system is just intuitively calculated as the
product of the one after each measurement, rather than derived from
the dynamical equation for the total evolution process. This treatment is appropriate for the cases with
ideal projective measurements. However, as we will show in Sec. III, in the
cases of general measurements which could be non-ideal, a more
first-principle analysis is required. There have also been some full
dynamical discussions of QZE and QAZE in the limit of continuous measurement
\cite{DQAZEhalf2006,DQAZE2005} (i.e., the cases where the system keeps
interacting with the measurement apparatus during the total evolution time,
or the time interval between two measurements are much smaller than the
duration time of each measurement) or the case of repeated measurements \cite%
{DQAZE2010}. However, in these discussions the measurements are performed via several
special physical systems, rather than general apparatus. To our
knowledge, so far there has not been a \textit{general} dynamical explaination
for QZE and QAZE of the unstable states in a quantum open system for arbitrary environment
spectrum, environment temperature and system-apparatus coupling.

In this paper, we provide a general dynamical approach for the QZE and
QAZE of unstable states in a quantum open system. For simplicity, we illustrate our central ideas with a
two-level system (TLS)
coupled to a heat bath of a multi-mode bosonic field. As pointed in Sec. VI,
it is straightforward to apply our model in open systems and
environments of other kinds.

In our approach, every
measurement process is described by the dynamical coupling between the
quantum open system and the apparatus. The postulation of wave function
collapse is not used. To describe the repeated measurements, we use a
multi-apparatus model that in each measurement, the open system interacts
with an individual apparatus \cite{DQAZEhalf2004}, while all the other
apparatus are left alone. As proved in Sec. II, regarding to the
evolution of the density operator of the quantum open system, this
multi-apparatus model of repeated measurements is equivalent with the
intuitive model where all the measurements are given by the same apparatus
and the state of this apparatus is initialized before each measurement. We
also point out, in our model there is no special restriction on the details
of the measurement process, e.g., the coupling between the to-be-measured system
 and the heat
bath. To be more practical, we concentrate on the case of ``
repeated measurements," i.e., the duration time of each measurement is much
smaller than the time interval between two measurements.

Using the multi-apparatus dynamical model of repeated measurements,
we completely investigate the effect of the measurements on both the short-time and long-time
evolution of the TLS,
and derive the QZE and QAZE in various cases. For the short-time evolution,
with time-dependent
perturbation theory, we obtain
 the short-time decay
rate of the excited state of the TLS, and show that when the TLS is under repeated ideal
projective measurements, the short-time decay rate given by our dynamical model is
exactly same as
the one from the wave packet collapse postulation \cite{QAZ}. In the case
of imperfect measurements, the corrections due to the non-ideality and the
finite duration time of the measurements can also be naturally obtained. We
prove that in the case of non-ideal measurements, QZE also occurs when the
measurements are frequent enough.

For the long-time evolution of the TLS, we derive the rate equation of the TLS by
calculating the complete long-time evolution of the TLS and the environment,
rather than by resetting the state of the environment after each
measurement. Since the Markovian approximation is not applied, our rate
equation can be used for the cases of heat baths with arbitrary correlation
time and the cases of measurements repeated with arbitrary frequency. The
rate equation clearly shows that the spectrum of the noise is effectively
tuned by the repeated measurements. Especially, we show that when
the measurements are frequent enough,
the time-local approximation and coarse-grained
approximation will be applicable. In this case, at zero temperature, the decay rate
given by the long-time rate equation is the same as the short-time
decay rate from our short-time perturbative calculation in Sec. IV.
 Furthermore, as a result of the counter-rotating terms in
 the coupling Hamiltonian of the TLS and the heat both, the decay rate of the ground state of the TLS
 may be varied to a  non-zero value by the periodic measurements, even at zero temperature.

Our model is also able to describe the QZE
or QAZE given by periodic phase modulation pulses~\cite%
{phaseM1,phaseM2,phaseM3} rather than measurements, since the former one can be
considered as a special kind of non-ideal measurement with complex decoherence factor. Our
calculation also provides a microscopic or full-quantum theory for the
recent proposals of stochastic control of quantum coherence~\cite%
{control1,control2,control3}, where the repeated projective measurements are
semi-classically described as a stochastic term in the Hamiltonian.

This paper is organized as follows. In Sec. II we post our
multi-apparatus dynamical model for the repeated quantum non-demolition measurements. In
Sec. III we show our dynamical description for the dissipative TLS under repeated
measurements. In Sec. IV we calculate the short-time decay rate of the excited state of the TLS
via time-dependent perturbation theory, and discuss the effects given by
non-ideal measurements or phase modulation pulses to QZE and QAZE. In Sec. V
we consider the long-time evolution of the system, derive the effective
noise spectrum experienced by the TLS and obtain the rate equation for the
long-time evolution at finite temperature. There are some conclusions and
discussions in Sec. VI.

\section{Multi-apparatuses for repeated measurements}

\begin{figure}[tbp]
\includegraphics[bb=41bp 182bp 570bp 523bp,clip,width=9cm]{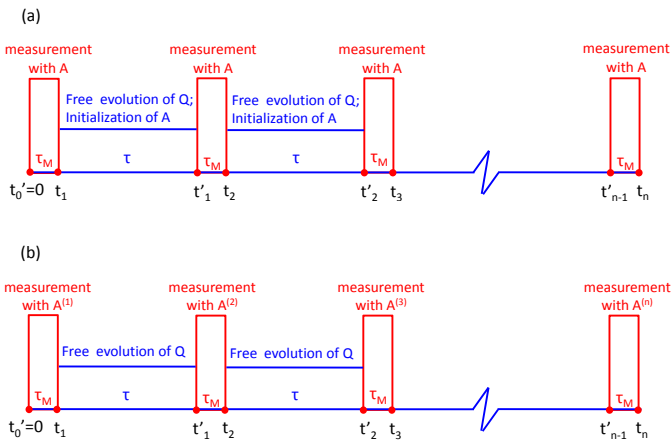}
\caption{(color online) (a) The single-apparatus model of repeated
measurements. All the measurements are performed with the same apparatus $A$%
. Before every measurement the state of $A$ is initialized to a given state $%
|\mathrm{app}\rangle $. (b) Multi-apparatus model for repeated measurements.
Each measurement is done with a individual apparatus with initial state
$|\mathrm{app}\rangle $.}
\end{figure}

In this paper, we discuss the quantum evolution of a dissipative TLS under
repeated quantum non-demolition measurements. There are two possible models
to describe the repeated quantum measurements, i.e., the single-apparatus
model and the multi-apparatus model. In the single-apparatus model, all the
measurements are completed via the same apparatus whose state is
``initialized" to a special one in the beginning of each measurement. In the
multi-apparatus model, each measurement is achieved by an individual
apparatus \cite{DQAZEhalf2004}. Namely, in every measurement, only one apparatus
interacts with the to-be-measured system while all the others are left
alone. In the recent experimental realization of QZE in cavity QED \cite%
{ExpCQED}, in every measurement the state of the cavity field is measured by
an individual ensemble of cold atoms. It can be considered as an illustration
of the multi-apparatus model.

In the current section, we prove that, regarding to the evolution of the
to-be-measured system (in our case, it is the TLS together with the heat
bath), the two models are equivalent. They can lead to the same evolution of
the density matrix of the to-be-measured system. For the convenience of our
calculation, in this paper we will use the multi-apparatus model in our
discussion for QZE and QAZE from the next section.

In the following we will first show the dynamical model of a single quantum
non-demolition measurement, and then do the formal calculation for the
evolution of the to-be-measured system under repeated measurements with
single-apparatus and multi-apparatus model. Our result shows that the density
matrix of the to-be-measured system has the same time evolution in both models.
For the generality of our discussion, in this section the
to-be-measured system is assumed to be a general multi-level quantum system.
From next section we will focus on the system of TLS together with the heat
bath.

\subsection{The dynamical model of a single quantum non-demolition
measurement}

According to the dynamical theory of quantum non-demolition measurements~%
\cite{QM}, the measurement process can be described by the coupling between
the to-be-measured system $Q$\ and the apparatus $A$. The total Hamiltonian
of $Q$\ and $A$\ has an expression of conditional dynamics
\begin{equation}
H_{M}=\sum_{j=1}|j\rangle \langle j|\otimes H_{j}
\end{equation}%
where $H_{j}$\ is the Hamiltonian of the apparatus $A$\ with respect to the $%
j$-th eigenstate $|j\rangle $\ of the observable of the open system. Before
the measurement, the system $Q$\ can be in any superposition state $%
\sum_{j=1}C_{j}|j\rangle $, while the apparatus is set in a pure state $%
|{\rm app}\rangle $. If the duration time of the measurement is $\tau _{\mathrm{M}}
$, the measurement leads to the transformation
\begin{equation}
\sum_{j=1}C_{j}|j\rangle |\mathrm{app}\rangle \rightarrow
\sum_{j=1}C_{j}|j\rangle |\mathrm{A}_{j}\rangle
\end{equation}%
where
\begin{equation}
|\mathrm{A}_{j}\rangle =e^{-iH_{j}\tau _{\mathrm{M}}}|\mathrm{app}\rangle
\label{adfdafd}
\end{equation}%
is the finial state of the apparatus with respect to the state $|j\rangle $\
of the system $Q$. Therefore, in the more general case, if the initial
density matrix of the system $Q$ is $\rho _{Q}$, after the measurement
the density matrix of $Q$ becomes
\begin{equation}
\mathcal{L}\left[ \rho _{Q}\right] \equiv \sum_{i,j}\langle i|\rho
_{Q}|j\rangle \langle \mathrm{A}_{i}|\mathrm{A}_{j}\rangle |i\rangle \langle
j|.  \label{sl}
\end{equation}%
Then the effect of the quantum non-demolition measurement on the
to-be-measured system $Q$\ can be described by the relevant decoherence
factors $\langle A_{i}|A_{j}\rangle $\ in the definition (\ref{sl}) of the
super-operator $\mathcal{L}$.

\subsection{The single-apparatus model of repeated measurements}

Now we consider the case of repeated measurements. As
shown in Fig. 1(a) and 1(b), we assume the measurement is performed once in
every time region $\left[ t_{n-1}^{\prime },t_{n}\right] $\ $(n=1,2,...;\
t_{0}^{\prime }=0)$\ with duration $\tau _{\mathrm{M}}$. During the
measurement, the Hamiltonian of the system $Q$\ and the relevant apparatus
is $H_{M}$. The length of the time intervals between every two neighbor
measurements is assumed to be $\tau $. The system evolves freely with
Hamiltonian $H_{Q}$\ in the time between the measurements.

In the single-apparatus model, all the measurements are performed via the
same apparatus $A$ (Fig. 1(a)). As we shown above, we assume the state of
the apparatus $A$\ is initialized to a given state $|{\rm app}\rangle $\ before
every measurement, and the information obtained by $A$\ from the last
measurement is ``erased". This initialization or erasing process can be done
via switching on the interaction between $A$\ and an external reservoir $R$,
like the spontaneous emission of the two-level atom. The similar technique
has also been used in our bang-bang cooling scheme for the nano-mechanical
resonator \cite{NAMR}.

Now we consider the evolution of the density matrix of $Q$\ in the
repeated-measurement process. At the initial time $t_{0}^{\prime }$, the
density matrix of $Q$\ and $A$\ is
\begin{equation}
\rho _{QA}\left( t_{0}^{\prime }\right) =\rho _{Q}\left( t_{0}^{\prime
}\right) |\mathrm{app}\rangle \langle \mathrm{app}|.
\end{equation}%
During the time region from $t_{0}^{\prime }$\ to $t_{1}$, the first measurement
is performed via the interaction between the system $Q$\ and the apparatus $A
$\ while the interaction between $A$\ and the reservoir $R$\ is switched
off. As shown above, at the ending time $t_{1}$\ of the first measurement,
the density matrix of the system $Q$ becomes%
\begin{equation}
\rho _{Q}\left( t_{1}\right) =\mathcal{L}\left[ \rho _{Q}\left( t'_{0}\right) %
\right]
\end{equation}%
where the super-operator ${\mathcal L}$ is defined in (\ref{sl}).

In the time region between $t_{1}$\ and $t_{1}^{\prime }$, the $Q$-$A$
interaction is switched off. The system $Q$\ experiences a free evolution
governed by the free Hamiltonian $H_{Q}$, while the
apparatus $A$\ and the reservoir $R$\ experiences a coupling which induces
the initialization of the state of $A$. We denote the evolution operator
given by the $A$-$R$\ coupling as $u_{AR}$. Then we have the total density
matrix $\rho _{T}$ of $Q,$ $A$ and $R$\ at time $t_{1}^{\prime }$\ as%
\begin{equation}
\rho _{T}\left( t_{1}^{\prime }\right) =u_{F}u_{AR}\rho _{T}\left(
t_{1}\right) u_{AR}^{\dagger }u_{F}^{\dagger }
\end{equation}%
with
\begin{equation}
u_{F}=\exp \left[ -iH_{Q}\tau \right] .
\end{equation}%
It is pointed out that, since $u_{F}$\ and $u_{AR}$\ are the operators for
different systems, we have%
\begin{equation}
\left[ u_{F},u_{AR}\right] =0.
\end{equation}%
Then we have the density matrix of the system $Q$\ at time $t_{1}^{\prime }$%
\begin{eqnarray}
\rho _{Q}\left( t_{1}^{\prime }\right)  &=&\mathrm{Tr}_{AR}\left[
u_{F}u_{AR}\rho _{T}\left( t_{1}\right) u_{AR}^{\dagger }u_{F}^{\dagger }%
\right]   \notag \\
&=&u_{F}\mathrm{Tr}_{AR}\left[ u_{AR}\rho _{T}\left( t_{1}\right)
u_{AR}^{\dagger }\right] u_{F}^{\dagger }  \notag \\
&=&u_{F}\rho _{Q}\left( t_{1}\right) u_{F}^{\dagger }  \notag \\
&=&\mathcal{UL}\left[ \rho _{Q}\left( t_{0}\right) \right]
\end{eqnarray}%
with the super-operator $\mathcal{U}$\ defined as $\mathcal{U}\left[ ...\right] =u_{F}\left[ ...%
\right] u_{F}^{\dagger }.$\ On the other hand, the density matrix of the
apparatus $A$\ at time $t_{1}^{\prime }$\ is given by%
\begin{eqnarray}
\rho _{A}\left( t_{1}^{\prime }\right)  &=&\mathrm{Tr}_{QR}\left[
u_{AR}u_{F}\rho _{T}\left( t_{1}\right) u_{F}^{\dagger }u_{AR}^{\dagger }%
\right]   \notag \\
&=&\mathrm{Tr}_{R}\left[ u_{AR}\mathrm{Tr}_{Q}\left[ u_{F}\rho _{T}\left(
t_{1}\right) u_{F}^{\dagger }\right] u_{AR}^{\dagger }\right]   \notag \\
&=&\mathrm{Tr}_{R}\left[ u_{AR}\rho _{AR}\left( t_{1}\right) u_{AR}^{\dagger
}\right]   \notag \\
&=&|\mathrm{app}\rangle \langle \mathrm{app}|
\end{eqnarray}%
with the density matrix $\rho _{AR}\left( t_{1}\right) $\ of $A$\ and $R$\
at time $t_{1}:$\
\begin{equation}
\rho _{AR}\left( t_{1}\right) =\mathrm{Tr}_{Q}\left[ \rho _{T}\left(
t_{1}\right) \right] .
\end{equation}%
In the last step, we have used the fact%
\begin{equation}
\mathrm{Tr}_{R}\left[ u_{AR}\rho _{AR}\left( t_{1}\right) u_{AR}^{\dagger }%
\right] =|\mathrm{app}\rangle \langle \mathrm{app}|.  \label{spl}
\end{equation}%
Namely, the $A$-$R$ coupling can make the density matrix of $A$\ to decay
to the unique steady state $|\rm{app}\rangle \langle \rm{app}|$, which is independent
on the density matrix of $A$ at the time $t_{1}$\ before the switching on of
the $A$-$R$\ coupling. That ``spontaneous-emission-like process" is the
physical explanation of the ``initialization" of the apparatus state. Eq. (%
\ref{spl}) is applicable when the influence of the $A$-$R$\ coupling on the
reservoir $R$ is negligible in the total evolution from time $t_{0}^{\prime }
$\ to $t_{N}^{\prime }$. Then the density matrix of $Q$\ and $A$\ at the
beginning time $t_{1}^{\prime }$\ of the second measurement becomes
\begin{equation}
\rho _{QA}\left( t_{1}^{\prime }\right) =\rho _{Q}\left( t_{1}^{\prime
}\right) |\mathrm{app}\rangle \langle \mathrm{app}|,
\end{equation}%
which has the same form as the one $\rho _{QA}\left( t_{0}\right) $\ at
the beginning of the first measurement. Therefore we can straightforwardly
generalize our above discussion to the time after $t_{1}^{\prime }$. Finally
we have the density matrix of the system $Q$\ at the time $t_{n}^{\prime }$\
after $n$\ measurements and $n$\ free evolutions:
\begin{equation}
\rho _{Q}\left( t_{n}^{\prime }\right) =\left( \mathcal{UL}\right) ^{n}\left[
\rho _{Q}\left( t_{0}\right) \right] .  \label{rrq}
\end{equation}

\subsection{ The multi-apparatus model of repeated
measurements}

Now we consider the multi-apparatus model for the repeated measurements. In
this model we assume there are many individual apparatus $A^{\left( 1\right)
},A^{\left( 2\right) },...$, each of which is in the same state $|{\rm app}\rangle $\
before the measurements. In the $m$-th measurement in the time region $\left[
t_{m-1}^{\prime },t_{m}\right] $, the system $Q$\ only interacts with the $m$%
-th apparatus $A^{\left( m\right) }$, and leave the other apparatus alone.
In the time interval between two measurements, the evolution of $Q$ is also
governed by the same Hamiltonian $H_{Q}$. Nevertheless, the initialization
of the state of the apparatus is not required in this multi-apparatus model
(Fig. 1(b)).

At the beginning time $t_{0}^{\prime }$\ of the first
measurement, the density matrix of $Q$\ and $A^{\left( 1\right) }$\ is
\begin{equation}
\rho _{QA^{\left( 1\right) }}\left( t_{0}^{\prime }\right) =\rho _{Q}\left(
t_{0}^{\prime} \right) |\mathrm{app}\rangle \langle \mathrm{app}|.  \label{rhoqa1}
\end{equation}%
At the time $t_{1}$, after the first measurement, the density matrix of the
system $Q$\ is also $\mathcal{L}\left[ \rho _{Q}\left( t_{0}^{\prime} \right) \right] $. At time
$t_{1}^{\prime }$, the density matrix of $Q$\ is
\begin{eqnarray}
\rho _{Q}\left( t_{1}^{\prime }\right) =u_{F}\rho _{Q}\left( t_{1}\right)
u_{F}^{\dagger}=\mathcal{UL}\left[ \rho _{Q}\left( t_{0}^{\prime} \right) \right] .
\end{eqnarray}%
Therefore, at the beginning time $t_{1}^{\prime }$\ of the second
measurement, the density matrix of $Q$\ and the apparatus $A^{\left(
2\right) }$\ is
\begin{equation}
\rho _{QA^{\left( 2\right) }}\left( t_{1}^{\prime }\right) =\rho _{Q}\left(
t_{1}\right) |\mathrm{app}\rangle \langle \mathrm{app}|  \label{rhoqa2}
\end{equation}%
with the same form as $\rho _{QA^{\left( 1\right) }}\left( t_{0}^{\prime
}\right) $. Then at time $t_{n}^{\prime }$ after the $n$-th measurements we
also have the density matrix of $Q$%
\begin{equation}
\rho _{Q}\left( t_{n}^{\prime }\right) =\left( \mathcal{UL}\right) ^{n}\left[
\rho _{Q}\left( t_{0}^{\prime} \right) \right].   \label{rrq1}
\end{equation}%
That is the same as the one (\ref{rrq}) given by the single-apparatus model.

Therefore, the single-apparatus and multi-apparatus model for the
repeated measurements can lead to the same evolution (\ref{rrq1}) of the
to-be-measured system $Q$. If we only consider the evolution of the system $Q$
under the repeated measurements, we can use either single-apparatus model or
multi-apparatus model in our calculation. In our discussion in the following
sections, the system $Q$\ includes the TLS and the heat bath which is
coupled to the TLS in the time region of free evolution. We will use the
multi-apparatus model to describe the repeated measurements periodically
performed on the TLS.

\section{Repeated Measurements about Two Level System}

\subsection{System and measurements}

In the above section we post our multi-apparatus model for the repeated
quantum measurements. From this section, we show our dynamical approach for the
QZE and QAZE of a dissipative TLS under repeated
measurements.

We consider a TLS coupled with a heat bath which is described as a
multi-mode bosonic field. The Hamiltonian of the total system is
\begin{eqnarray}
H_{F} &=&\omega _{eg}|e\rangle ^{(S)}\langle e|+\sum_{\vec{k}}\omega _{\vec{k%
}}a_{\vec{k}}^{\dag }a_{\vec{k}}  \notag \\
&&+|e\rangle ^{\left( S\right) }\langle g|\sum_{\vec{k}}\left( g_{\vec{k}}a_{%
\vec{k}}+g_{\vec{k}}^{\ast }a_{\vec{k}}^{\dagger }\right) +h.c.  \label{hf}
\end{eqnarray}%
Here $|g\rangle ^{\left( S\right) }$ and $|e\rangle ^{\left( S\right) }$ are
the ground and excited states of the TLS, $a_{\vec{k}}^{\dag }$ and $a_{\vec{%
k}}$ are respectively the creation and annihilation operators of the boson
in the {$\vec{k}$-th heat-bath mode with frequency $\omega _{\vec{k}}$},
 while $g_{\vec{k}}$ is the relevant coupling intensity between the boson
and the TLS. In our Hamiltonian $H_{F}$, the rotating wave approximation is
not used  so that the possible
effects given by the counter-rotating terms can be included \cite{Hang,Qing}.

In this paper we consider the evolution of such a dissipative TLS
under repeated measurements. As shown in the above section,
we use the multi-apparatus model to
describe the repeated measurements. We also assume the $n$-th measurement is
performed in the time region $\left[ t_{n-1}^{\prime },t_{n}\right] $
$(n=1,2,...;\ t_{0}^{\prime }=0)$ with the duration time $\tau _{\mathrm{M}}$
(Fig. 2(a) and Fig. 2(b)). The time intervals between two neighbor
measurements are also assumed to have the same length $\tau $.
In these time intervals between the measurements, the system evolves
freely {under} Hamiltonian $H_{F}$. In
the multi-apparatus model, we assume there are many {apparatuses} $A^{\left(
1\right) },...,A^{\left( n\right) },...$ which can be individually coupled
with the TLS and distinguish between the states $|e\rangle ^{\left( S\right)
}$ and $|g\rangle ^{\left( S\right) }$. In this paper we denote $|\rangle
^{\left( S\right) }$ for the quantum state of the TLS, $|\rangle
^{\left( B\right) }$ for the state of the heat bath, $|\rangle ^{\left(
n\right) }$ for the state of the $n$-th apparatus $A^{\left( n\right) }$ and
$|\rangle ^{\left( A\right) }$ for the state of all the apparatus. Before
the measurement, every apparatus is {\ initially} in a pure state. During
the $n$-th measurement, the TLS is coupled with the $n$-th apparatus $%
A^{\left( n\right) }$ and decoupled with all the other apparatuses $%
A^{\left( m\neq n\right) }$.

\begin{figure}[tbp]
\includegraphics[bb=44bp 66bp 539bp 768bp,clip,width=8.5cm]{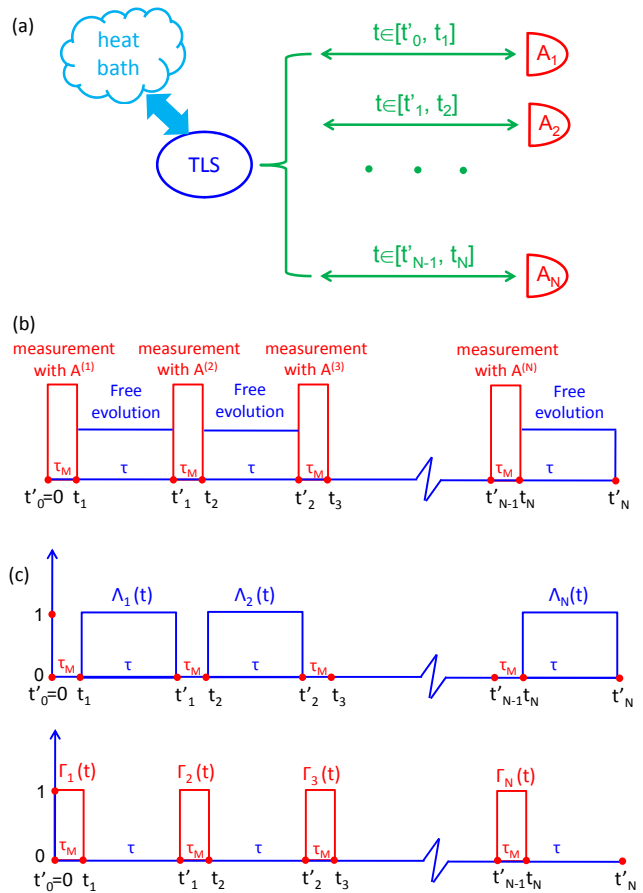}
\caption{(color online) (a) The multi-apparatus model of the
 repeated quantum measurements of a dissipative TLS. In
each measurement, the TLS is coupled with an individual apparatus and
leaves other apparatus alone. The $n$-th measurement occurs in the time
region between $t_{n-1}^{\prime }$ and $t_{n}$. The TLS is also coupled with
the heat bath. (b) The time sequence of repeated measurements and free
evolution governed by $H_{F}$ defined in (\ref{hf}). (c) The
definitions of the functions {$\Gamma _{n}(t)$ and $\Lambda _{n}(t)$.} }
\end{figure}

As shown in Sec. II, in the quantum non-demolition measurements
the coupling between the system and the apparatus $A^{\left( n\right) }$ has
an expression of conditional dynamics:
\begin{eqnarray}
H_{M}^{\left( n\right) }=|e\rangle ^{\left( S\right) }\langle e|\otimes
H_{e}^{\left( n\right) }+|g\rangle ^{\left( S\right) }\langle g|\otimes
H_{g}^{\left( n\right) },
\end{eqnarray}%
where $H_{e}^{\left( n\right) }$ $(H_{g}^{\left( n\right) })$ is the
Hamiltonian of the apparatus $A^{\left( n\right) }$ with respect to the
state $|e\rangle ^{\left( S\right) }$ $\left( |g\rangle ^{\left( S\right)
}\right) $ of the TLS. In this paper, we assume the duration time $\tau _{%
\mathrm{M}}$ of the measurement is so small that the interaction between the
TLS and the heat bath and the decay of the excited state can be neglected
during the measurement.

In the $n$-th measurement, if the state of the TLS before the measurement is
$\alpha |e\rangle ^{\left( S\right) }+\beta |g\rangle ^{\left( S\right) }$,
and the one of the apparatus $A_{n}$ is $|\mathrm{app}\rangle ^{\left(
n\right) }$, then the transformation given by the measurement can be
described as
\begin{eqnarray*}
&&\left( \alpha |e\rangle ^{\left( S\right) }+\beta |g\rangle ^{\left(
S\right) }\right) |\mathrm{app}\rangle ^{\left( n\right) } \\
&\rightarrow &\alpha e^{-i\omega _{eg}\tau _{\mathrm{M}}}|e\rangle ^{\left(
S\right) }|A_{e}\rangle ^{\left( n\right) }+\beta |g\rangle ^{\left(
S\right) }|A_{g}\rangle ^{\left( n\right) }
\end{eqnarray*}%
with $|A_{g}\rangle ^{\left( n\right) }$ and $|A_{e}\rangle ^{\left(
n\right) }$ the states of the apparatus attached to the ground and excited
states of the TLS
\begin{eqnarray}
|A_{g,e}\rangle ^{\left( n\right) }=e^{-iH_{g,e}^{\left( n\right) }\tau _{%
\mathrm{M}}}|\mathrm{app}\rangle ^{\left( n\right) }.
\end{eqnarray}%
The ideality of the measurement is described by the overlap of the states $%
|A_{g}\rangle ^{\left( n\right) }$ and $|A_{e}\rangle ^{\left( n\right) }$
or the decoherence factor%
\begin{eqnarray}
\gamma _{n}e^{i\theta _{n}}=\ ^{\left( n\right) }\langle A_{e}|A_{g}\rangle
^{\left( n\right) };\ \gamma _{n},\theta _{n}\in \mathrm{Reals}.\label{gn}
\end{eqnarray}%
In the case of ideal projective measurement, these two states are orthogonal {%
to} 
each other and we have $\gamma _{n}=0$.

\subsection{The total Hamiltonian in the interaction picture}

Under our above assumptions, the quantum evolution of the total system
including the TLS, the heat bath and the {apparatuses} 
is governed by the time-dependent Hamiltonian
\begin{eqnarray}
&&H\left( t\right) =\omega _{eg}|e\rangle ^{(S)}\langle e|+\sum_{\vec{k}%
}\omega _{\vec{k}}a_{\vec{k}}^{\dag }a_{\vec{k}}+\sum_{n}H_{M}^{\left(
n\right) }\Gamma _{n}\left( t\right)  \notag \\
&&+\left( |e\rangle ^{\left( S\right) }\langle g|+
|g\rangle ^{\left( S\right) }\langle e|\right)\left(\sum_{\vec{k}}g_{\vec{k}}a_{%
\vec{k}}+h.c.\right) \sum_{n}\Lambda _{n}\left( t\right), \nonumber\\
&& \label{htotal}
\end{eqnarray}
where the functions $\Gamma _{n}\left( t\right) $ and $\Lambda _{n}\left(
t\right) $ are defined as in {Fig.~2(c):} 
\begin{eqnarray}
\Gamma _{n}\left( t\right) =\left\{
\begin{array}{c}
1;t\in \left[ t_{n-1}^{\prime },t_{n}\right] \\
0;\ \mathrm{otherwise}%
\end{array}%
\right. ;
\end{eqnarray}%
and
\begin{eqnarray}
\Lambda _{n}\left( t\right) =\left\{
\begin{array}{c}
1;t\in \left[ t_{n},t_{n}^{\prime }\right] \\
0;\ \mathrm{otherwise}%
\end{array}%
\right. .
\end{eqnarray}

To solve the quantum evolution of our system, we use the interaction picture
where the quantum state $|\Psi \left( t\right) \rangle _{I}$ is defined as%
\begin{eqnarray}
|\Psi \left( t\right) \rangle _{I}=\exp \left[ i\int_{0}^{t}H_{0}\left(
t^{\prime }\right) dt^{\prime }\right] |\Psi \left( t\right) \rangle .
\end{eqnarray}%
Here $|\Psi \left( t\right) \rangle $ is the state of the total system in
the Schr\"{o}dinger picture and $H_{0}$ is given by
\begin{eqnarray}
H_{0}\left( t\right) =\omega _{eg}|e\rangle ^{(S)}\langle e|+\sum_{\vec{k}%
}\omega _{\vec{k}}a_{\vec{k}}^{\dag }a_{\vec{k}}+\sum_{n}H_{M}^{\left(
n\right) }\Gamma _{n}\left( t\right) .  \label{h0}
\end{eqnarray}%
In the interaction picture, the quantum state $|\Psi \left( t\right) \rangle
_{I}$ satisfies the Schr\"{o}dinger equation%
\begin{eqnarray}
i\frac{d}{dt}|\Psi \left( t\right) \rangle _{I}=H_{I}\left( t\right) |\Psi
\left( t\right) \rangle _{I}  \label{she}
\end{eqnarray}%
with the Hamiltonian $H_{I}\left( t\right) $ given by%
\begin{eqnarray}
H_{I}\left( t\right) =|e\rangle ^{\left( S\right) }\langle g|\hat{f}%
_{B}\left( t\right) \hat{f}_{A}\left( t\right) +h.c.,  \label{hit}
\end{eqnarray}%
where the operators $\hat{f}_{B}\left( t\right) $ and $\hat{f}_{A}\left(
t\right) $ are defined as
\begin{eqnarray}
\hat{f}_{B}\left( t\right) &=&\sum_{\vec{k}}\left(g_{\vec{k}}a_{\vec{k}%
}e^{-i\Delta_{\vec k} t}+g_{\vec{k}}^{\ast }a_{\vec{k}}^{\dag }e^{i\left(
\omega _{eg}+\omega _{\vec{k}}\right) t}\right) ,  \label{fa} \\
\hat{f}_{A}\left( t\right) &=&\sum_{n=1}\Lambda _{n}\left( t\right) M_{n}.
\label{fb}
\end{eqnarray}%
Here the detuning $\Delta _{\vec{k}}$ takes the form
\begin{eqnarray}
\Delta _{\vec{k}}=\omega _{\vec{k}}-\omega _{eg}
\end{eqnarray}%
and the unitary operator $M_{n}$ is given by%
\begin{eqnarray}
M_{n}=\prod_{l=1}^{n}\exp \left[ iH_{e}^{\left( l\right) }\tau _{\mathrm{M}}%
\right] \exp \left[ -iH_{g}^{\left( l\right) }\tau _{\mathrm{M}}\right] .
\label{mn}
\end{eqnarray}

With the help of the interaction picture, the effect of the measurements {is}
packaged in the definition of the operator $M_n$. As shown below, in this
interaction picture our calculations are significantly simplified and we can
express all the effects from the measurements in terms of the decoherence
factor $\gamma_ne^{i\theta_n}$ defined in {Eq.~(\ref{gn}).} In the following
two sections we will derive the short-time and long-time {\ evolutions}
of the TLS with the calculations in this interaction picture.

\section{Short-time evolution: first-order perturbation theory}

In this section we calculate the short-time decay rate of the excited state $%
|e\rangle ^{\left( S\right) }$ of the TLS under repeated measurements. For simplicity, we
only consider the zero-temperature case where the
initial state of our system at $t=t^{\prime}_0=0$ as
\begin{eqnarray}
|\Psi \left( 0\right) \rangle _{I}=|e\rangle ^{\left( S\right) }|\mathrm{vac}
\rangle ^{\left( B\right) }\prod_{m}|\mathrm{app}\rangle ^{\left( m\right) },
\label{psi0}
\end{eqnarray}%
where $|\mathrm{vac}\rangle ^{\left( B\right) }$ is the vacuum state of the
bosonic field and $|\mathrm{app}\rangle ^{\left( m\right) }$ the initial
state of the $m$-th apparatus. We consider the evolution of the system from $%
t=0$ to a finial time $t=t_F$, which, for simplicity is assumed to be an
integer multiple of the period $\tau+\tau_{\mathrm{M}}$ of the measurements.

The finial state $|\Psi \left( t_{F}\right) \rangle _{I}$ of the total system
is the solution of the Schr\"{o}dinger equation (\ref{she}). It can be expressed
as a functional series of $H_I(t)$:
\begin{eqnarray}
|\Psi \left( t_{F}\right) \rangle _{I}=\left( 1-i\int_{0}^{t_{F}}H_{I}\left(
t\right) dt+...\right) |\Psi \left( 0\right) \rangle _{I}.
\end{eqnarray}%
The survival probability $P_{e}\left( t_{F}\right) $ of the state $%
|e\rangle ^{\left( S\right) }$ is given by
\begin{eqnarray}
P_{e}\left( t_{F}\right) &=&1-\ _{I}\langle \Psi \left( t_{F}\right)
|(|g\rangle ^{\left( S\right) }\langle g|)|\Psi \left( t_{F}\right) \rangle
_{I}.\label{sp}
\end{eqnarray}
We can define the short-time decay rate $R(\tau,t_F)$ of
$|e\rangle ^{\left( S\right) }$ as
\begin{eqnarray}
P_{e}\left( t_{F}\right)=1-R(\tau,t_F) t_{F}.
\end{eqnarray}%
Therefore, $R\left( \tau,t_F \right)$ can also be expressed as a
functional series of $H_I(t)$. When the total evolution time $t_F$ is small enough,
we can only keep the lowest-order term (in our problem it is the second-order term)
of $H_I(t)$ in $R(\tau,t_F)$. Then we have the short-time decay rate
\begin{eqnarray}
&&R(\tau,t_F)\approx\frac{\tau ^{2}}{t_{F}}\sum_{\vec{k}}|g_{\vec{k}}|^{2}\mathrm{sinc}%
^{2}\left( \Delta _{\vec{k}}\tau /2\right)  \notag  \label{rtau} \\
&&\times \left[ N+2\sum_{m=1}^{N}\sum_{n=1}^{m-1}\mathrm{Re}\left( e^{i\left(
n-m\right) \Delta _{\vec k }\left( \tau +\tau _{\mathrm{M}}\right)
}{}\prod_{l=n+1}^{m}\gamma _{l}e^{i\theta _{l}}\right) \right] ,  \notag \\
&& \label{bigr}
\end{eqnarray}%
where
\begin{eqnarray}
N=\frac{t_{F}}{\tau +\tau _{\mathrm{M}}}.
\end{eqnarray}

In the case of ideal projective measurements with negligible duration time
or $\tau _{\mathrm{M}}=0$, we have $\gamma _{l}=0$ and then the short-time decay rate $R(\tau,t_F)$
in Eq. (\ref{rtau}) becomes a $t_F$-independent one:%
\begin{eqnarray}
R(\tau,t_F)=R_{\mathrm{pro}}\left( \tau \right) =\tau \sum_{\vec{k}}|g_{\vec{k}}|^{2}
\mathrm{sinc}^{2}\left( \Delta _{\vec{k}}\tau /2\right) ,  \label{rpro}
\end{eqnarray}%
which is the result given by Ref.~\cite{QAZ} with the postulation of wave
function collapse. Therefore, the QZE or QAZE based on the ideal decay rate $%
R_{\mathrm{pro}}\left( \tau \right) $ can also be obtained with our model.

In the case of periodic non-ideal measurements, if the common upper limit of
the modului $\gamma _{n}$
of the decoherence factors in the measurements are smaller than unit, i.e., we
$\gamma _{n}\leq\gamma _{\max }<1$
for any $n$, then Eq.~(\ref{rtau}) gives
\begin{eqnarray}
R(\tau,t_F)\leq \tau \sum_{\vec{k}}|g_{\vec{k}}|^{2}\left( 1+\frac{2\gamma _{\max }}{%
1-\gamma _{\max }}\right) \mathrm{sinc}^{2}\left( \frac{\Delta _{\vec{k}%
}\tau }{2}\right).  \label{tau0}
\end{eqnarray}%
Therefore, for a fixed value of the total evolution time $t_F$,
we have
\begin{eqnarray}
\lim_{\tau \rightarrow 0}R(\tau,t_F)=0.
\end{eqnarray}%
It means that, in the cases of non-ideal measurements with $\gamma _{\max }<1$,
the QZE also occurs in the limit that the measurement are frequent enough.

In Refs.~\cite{DQAZEhalf2001,DQAZEhalf2001b,DQAZEhalf2004,DQAZEhalf2006},
the authors have calculated the total survival probability $P_{e}\left(
t_{F}\right) $ {as} 
the product of the one after each measurement, i.e., the relationship
\begin{eqnarray}
P(t_F)=P(\tau)^N \label{pro}
\end{eqnarray}
is assumed.
 This simplification yields
that the short-time decay rate $R(\tau,t_F)$ is the one in a single period of
measurement, or
\begin{eqnarray}
R(\tau,t_F)=R(\tau,\tau).  \label{pror}
\end{eqnarray}
In our case {of} 
$\tau \gg \tau _{\mathrm{M}}$, Eq. (\ref{pror}) leads to $R(\tau,t_F)=R_{\mathrm{%
pro}}\left( \tau \right) $. Therefore, the intuitive treatments
(\ref{pro}) and (\ref{pror}) are reasonable in the cases of ideal
measurements with $\gamma_n =0$, and required to be improved to Eq.~(\ref{bigr}%
) for the cases of 
non-ideal measurements and non-zero $\gamma_n$.

\subsection{The effect given by non-ideal measurements}

To further explore the physical meaning of the short-time
 decay rate given by Eq.~(\ref%
{rtau}), we consider the simple case with identical non-ideal measurements.
In this case the decoherence factor $\gamma _{n}e^{i\theta _{n}}$ takes $n$%
-independent value $\gamma e^{i\theta }$ ($\gamma <1$). In the large $N$
limit with $N\gg1/(1-\gamma )$, Eq.~(\ref{rtau}) gives a $t_F$-independent
short-time decay rate:
\begin{eqnarray}
R(\tau,t_F)&\approx &R_{\mathrm{mea}}\left( \tau \right)\nonumber\\
&\equiv &\int d\eta G\left( \frac{\eta }{\tau }%
+\omega _{eg}\right)h\left(\gamma ,\theta -\eta \right)\mathrm{sinc}^{2}
\frac{\eta }{2}. \ \  \label{rmea2}
\end{eqnarray}
Here we have used the approximation $\tau /(\tau +\tau _{\mathrm{M}})\approx
1$. In the above expression the spectrum function $G\left( \omega \right) $
of the heat bath defined as
\begin{eqnarray}
G(\omega )=\sum_{\vec{k}}\left\vert g_{\vec{k}}\right\vert ^{2}\delta
(\omega -\omega _{\vec{k}}) \label{g}
\end{eqnarray}%
and the function {$h\left( \gamma ,x\right) $} 
is given by
\begin{eqnarray}
h\left( \gamma ,x\right) =\frac{1-\gamma ^{2}}{1+\gamma ^{2}-2\gamma \cos x}.
\label{h}
\end{eqnarray}%
It is obvious that, in the case $\gamma =0$ we have {$h\left( 0 ,x\right) =1$%
} 
and the decay rate $R_{\mathrm{mea}}\left( \tau \right) $ in Eq.~(\ref{rmea2}%
) returns to the result $R_{\mathrm{pro}}\left( \tau \right) $ given by
ideal projective measurements. In {the case} 
of repeated non-ideal measurements, the function $h\left( \gamma ,\theta
-\eta \right) $ in the right hand side of Eq.~(\ref{rmea2}) would tune the
shape of the function to be integrated and then change the value of $R_{%
\mathrm{mea}}\left( \tau \right) $.

\begin{figure}[tbp]
\includegraphics[bb=99bp 263bp 521bp 566bp,clip,width=8cm]{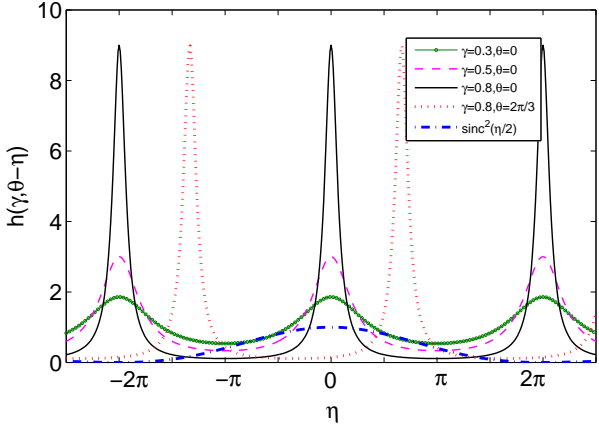}
\caption{(color online) The function $h(\gamma,\theta-%
\eta)$ defined in Eq.~(\ref{h}). Here we plot the curves
with $\gamma=0.3,\ \theta=0$ (green line with open circle); $%
\gamma=0.5,\ \theta=0$ (pink dashed line); $\gamma%
=0.8,\ \theta=0$ (black solid line) and $\gamma=0.8,\
\theta=2\pi/3$ (red dotted line). It is clearly shown that
the position of the peak of $h(\gamma,\theta-\eta)$
is determined by the value of $\theta$, while the width of the peak
decreases when $\gamma$ is increased. As a comparison, we also plot
the function $\mathrm{{sinc}^2(\eta/2)}$ (blue dashed dotted line).}
\end{figure}

In {Fig.~3} 
we plot the function $h\left( \gamma ,\theta -\eta \right) $ with respect to
different values of decoherence factor $\gamma e^{i\theta }$. It is easy to
prove that $h\left( \gamma ,\theta -\eta \right) $ takes peak value at the
points $\eta =\theta +2n\pi $ ($n=0,\pm 1$,....) with {the} width of the
order of $\arccos \gamma $. Therefore, the effects given by $h\left( \gamma
,\theta -\eta \right) $ on the decay rate $R_{\mathrm{mea} }\left( \tau
\right) $ in Eq.~(\ref{rmea2}) seriously depend on the values of both $%
\gamma $ and $\theta $.

To illustrate these effects, we calculate the decay rate $R_{\mathrm{mea}%
}\left( \tau \right) $ of a TLS in an environment with noise spectrum%
\begin{eqnarray}
G(\omega )=G_{H}(\omega )\equiv\left\{
\begin{array}{c}
\frac{\omega }{\left( 1+\left( \frac{\omega }{\omega _{c}}\right)
^{2}\right) ^{4}};\omega\geq 0 \\
\ 0;\ \ \ \ \ \ \ \ \ \ \ \omega<0%
\end{array}%
\right. ,  \label{gh}
\end{eqnarray}%
which is, up to a global factor, the same as the noise spectrum of the 2p-1s
transition of the hydrogen atom \cite{hyd,hyd2}. As in the case of realistic
hydrogen atom, here we also take $\omega _{c}=549.5\omega _{eg}$. The decay
rates $R_{\mathrm{mea}}\left( \tau \right) $ with $\theta =0,\pi /2,\pi
,3\pi /2$ and various values of $\gamma $ are shown in {Fig.~4.} 

Our results {show} 
that, with any values of $\gamma $ and $\theta $, the short-time decay rate $R_{\mathrm{%
mea}}\left( \tau \right) $ approaches to zero in the limit $\tau \rightarrow
0$. That is, the QZE always occurs when the measurements are repeated
frequently enough. This is consistent with the conclusion in Eq.~(\ref{tau0}%
). On the other hand, in the limit $\tau \rightarrow \infty $, $R_{\mathrm{%
mea}}\left( \tau \right) $ approaches to the natural decay rate $R_{\mathrm{%
GR}}$ given by the Fermi golden rule:%
\begin{eqnarray}
R_{\mathrm{GR}}=2\pi G\left( \omega _{eg}\right) .  \label{rgr}
\end{eqnarray}%
Therefore, the quantum Zeno and anti-Zeno effects only occur 
when $\tau $ is small enough.

When the phase $\theta =0$, as shown in {Fig.~4(a),} 
both {the} QZE and QAZE can also appear even when the decoherence factor $%
\gamma $ is nonzero. However, when the value of $\gamma $ or non-ideality of
the measurements is increased, the region of $\tau $ for the appearance of
QZE and QAZE becomes narrower. That is, in the case of large $\tau$, the QZE
and QAZE with large $\gamma$ {are} 
always less significant than the ones with small $\gamma$. In this sense the
two effects are suppressed by the non-ideal measurements with real decoherence
factors.

For the cases {of} 
non-zero phase $\theta $, the effects given by the non-ideal measurements
are from both the non-zero value of $\gamma$ and the repeated phase
modulation given by the phase factor $e^{i\theta}$. As shown in {\
Figs.~4(b)-4(d),} 
the final behavior of $R_{\mathrm{mea}}$ sensitively dependents on $\theta$%
. Especially, with specific values of $\theta $ and $\gamma $, the non-ideal
measurements can enhance either QZE or QAZE. Both the $\tau$-region where
the QZE occurs and the peak value of the decay rate in the QAZE can be
enlarged by the complex value of the decoherence factor $\gamma e^{i\theta}$.
Finally, the total region of $\tau $ for the
occurrence of QZE and QAZE with non-zero $\gamma $ is possible to be the same as the
one for the cases of $\gamma =0$, or even larger than the latter. That is
due to the complicated behavior of the function $h\left( \gamma ,\theta
-\eta \right) $.

\begin{figure}[tbp]
\includegraphics[bb=47bp 183bp 510bp 648bp,clip,width=8.5cm]{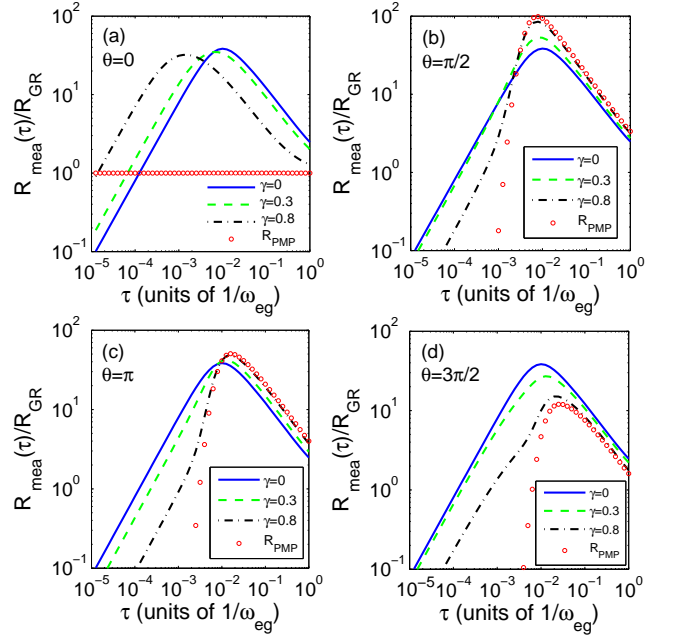}
\caption{(color online) The short-time decay rate $R_{\mathrm{mea}}(\tau)$
defined in Eq.~(\ref{rmea2}) of a dissipative TLS under repeated
measurements. Here we take the noise spectrum in Eq.~(\ref{gh}) and
illustrate the cases with $\theta=0$ (a), $\theta=\pi%
/2$ (b), $\theta=\pi$ (c), $\theta=3\pi/2$
(d) and $\gamma=0$ (blue solid line), $\gamma=0.3$ (green
dashed line), $\gamma=0.8$ (black dashed-dotted line). We also plot $%
R_{\mathrm{mea}}(\tau)$ (red open circle) defined in Eq.~(
\ref{rp}) given by frequent phase modulation pulses with approximation (%
\ref{gd}). }
\end{figure}

\subsection{The quantum Zeno and anti-Zeno effect given by phase modulation
pulses}

As shown in the above subsection, in the cases of non-ideal measurements
with $\gamma \neq 0$, both the QZE and QAZE are possible to be enhanced in the
cases with nonzero phase shift $\theta $ in each measurement. Actually in
the most extreme cases of $\gamma =1$, the two effects can also appear with
nonzero phase $\theta $. In that case, the measurements reduce to the
phase modulation pulses which can induce periodical jumps for the relative
phase between $|e\rangle ^{(S)}$ and $|g\rangle ^{(S)} $. The QZE and QAZE
given by the phase modulation pulses have been discussed in Ref.~\cite%
{phaseM2,phaseM3} in detail. Here we show that our dynamical model is also
able to describe these effects.

If the pulse is repeatedly performed with period $\tau $ and negligible
duration time $\tau _{\mathrm{M}}$, then the short-time decay rate in Eq.~(\ref{bigr})
becomes
\begin{eqnarray}
R(\tau,t_F)&=&R_{\mathrm{PMP}}(\tau)\nonumber\\
&\equiv&\int d\eta G\left( \frac{\eta }{\tau }+\omega _{eg}\right)
g\left( \theta -\eta \right) \mathrm{sinc}^{2}\left( \frac{\eta }{2}\right),\ \
\label{rp}
\end{eqnarray}%
where the function $g(x)$ is defined as
\begin{eqnarray}
g\left( x\right) =\frac{1}{N}\frac{\sin ^{2}\left( Nx/2\right) }{\sin
^{2}\left( x/2\right) }.
\end{eqnarray}%
This result is essentially equivalent to the one in Refs.~\cite%
{phaseM2,phaseM3}.

From now on we assume the function $G(\omega+\omega_{eg})$ has a finite
width $\Delta\omega$, i.e., $G(\omega+\omega_{eg})$ takes nonzero value only
in the region $-\Delta\omega<\omega<\Delta\omega$. Under this assumption,
when the evolution time $t_{F}$ is large enough so that $2\pi /t_{F}$ is
much smaller than the width $\Delta\omega $, we have
\begin{eqnarray}
g(x)\approx 2\pi \sum_{n=-\infty }^{\infty }\delta \left( x+2n\pi \right).
\label{gd}
\end{eqnarray}%
If the period $\tau $ of the pulses is short enough so that $\tau <\pi
/\Delta \omega $ and
\begin{eqnarray}
G(-\pi/\tau+\omega_{eg})=G(\pi/\tau+\omega_{eg})\simeq0,  \label{g0}
\end{eqnarray}
the function $G(\eta/\tau+\omega_{eg})$ is localized in the region $%
-\pi<\eta<\pi$. Then the simplification (\ref{gd}) implies
\begin{eqnarray}
R_{\mathrm{PMP}} &&\approx 2\pi G\left( \frac{\theta }{\tau }%
+\omega_{eg}\right) \mathrm{sinc}^{2}\left( \frac{\theta }{2}\right)  \notag
\\
&&+2\pi G\left( \frac{\theta }{\tau }-\frac{2\pi }{\tau }+\omega_{eg}\right)
\mathrm{sinc}^{2}\left( \frac{\theta }{2}-\pi \right),
\end{eqnarray}%
where we have assumed $\theta \in \left[ 0,2\pi \right] $. In this case, as
a result of Eq.~(\ref{g0}), one can further find some special $\tilde{\theta}
$ so that $\tilde{\theta}/\tau-2\pi/\tau <-\Delta \omega $ and $\tilde{\theta%
}/\tau >\Delta \omega $, which makes $G(\tilde{\theta}/\tau +\omega_{eg})
\sim G(\tilde{\theta}/\tau-2\pi/\tau+\omega_{eg}) \sim 0$ or $R_{\mathrm{PMP}%
}\sim 0$. There are also other special angles $\theta ^{\ast }$ which
makes $G(\theta ^{\ast }/\tau+\omega_{eg}) $ or $G(\theta ^{\ast }/\tau-2\pi
/\tau+\omega_{eg}) $ take the maximum value $G^{\max }$ of $G\left( \omega
\right) $. Therefore, the decay rate $R_{\mathrm{PMP}}$ can be tuned in the
broad region between $0$ and some maximum value which is of the order of $2\pi
G^{\max }$. This tuning effect is also predicted in Refs.~\cite%
{phaseM2,phaseM3}.

We also point out that, in the limit $\gamma \rightarrow 1$ the function $%
h\left( \gamma ,x\right) $ in the last subsection has the same behavior%
\begin{eqnarray}
h\left( \gamma\rightarrow 1,x\right) \approx 2\pi \sum_{n=-\infty }^{\infty
}\delta \left( x+2n\pi \right)
\end{eqnarray}%
as $g(x)$ in the large $N$ limit. Therefore the results in Eq.~(\ref{rp})
for $2\pi /t_{F}<<\Delta \omega $ can also be obtained from Eq.~(\ref{rmea2}%
).

In {Fig.~4} 
we also plot the short-time decay rate $R_{\mathrm{PMP}}$ under approximation (\ref{gd}%
) with the spectrum (\ref{gh}) and different values of the angle $\theta$.
It is shown that, when $\gamma$ is close to unit the short-time decay rate $R_{\mathrm{%
mea}}$ in Eq.~(\ref{rmea2}) is quite close to $R_{\mathrm{PMP}}$. In
this case the behavior of the short-time decay rate is dominated by the repeated phase
modulation. 

\section{The Long-Time Evolution: Rate Equation}

In the above section we have considered the short-time evolution of the dissipative
TLS under repeated measurements or phase modulation pulses. We obtain the short-time
decay rates via perturbative calculation based on our pure dynamical model of repeated measurements.
The perturbative approach is simple and straightforward. Nevertheless, the results are only applicable
when the total evolution time $t_F$ is short.

In this section, we go beyond the short-time calculation and consider the long-time evolution of the TLS under
repeated measurements. The problem with projective measurements {has} 
been considered in Refs.~\cite{control1,control2,control3} in a
semi-classical approach with measurements described by a stochastic term in
the Hamiltonian. Here we provide a full-quantum theory which can be used for
the cases of either ideal or non-ideal measurements. The previous results
\cite{phaseM2,phaseM3} on the long-time evolution of a dissipative TLS under
repeated phase modulation pulses can also be derived in our approach.

For simplicity, here we assume the measurements are identical with the
decoherence factor $\gamma e^{i\theta }$. We first deduce the general form
of the rate equation of the TLS in terms of the effective time-correlation
function of the environment, and then derive the simplified form of the rate
equation under the time-local and coarse-grained approximation.

\subsection{The general rate equation and effective time-correlation
function of the environment}

To derive the rate equation for the TLS, we first consider the TLS and {%
the apparatuses} 
as a total system interacting with the heat bath. The evolution of the
density matrix $\rho ^{\left( SA\right) }$ of such a combined system can be
described by master equation given by Born
approximation (Eq.~(9.26) of Ref.~\cite{opensystem}):
\begin{eqnarray}
&&\frac{d}{dt}\rho^{\left( SA\right)}\left( t\right)\nonumber\\
&=&-\int_{0}^{t}ds\mathrm{%
Tr}_{B}\left[ H_{I}\left( t\right) ,\left[ H_{I}(s),\rho^{(SA)}(s)\otimes\rho ^{(B)}%
\right]\right],\ \ \ \label{me1}
\end{eqnarray}
where $H_{I}\left( t\right) $ is defined in Eq.~(\ref{hit}) and $\rho
^{\left( B\right) }$ is the initial density matrix of the heat bath, which
is assumed to be in the thermal equilibrium state at temperature $T$. It is
pointed out that, in Eq.~(\ref{me1}) we do not perform the Markovian
approximation.

The evolution equation for the density matrix $\rho ^{\left( S\right)
}\left( t\right) $ of the TLS can be obtained by tracing out the states of
the apparatuses in Eq.~(\ref{me1}). According to Eq.~(\ref{hit}), in the
calculation we need to evaluate the values of
\begin{eqnarray}
&&\mathrm{Tr}_{A}\left[ \hat{f}_{A}\left( t\right) \hat{f}_{A}^{\dag }\left(
s\right) \rho ^{\left( SA\right) }\left( s\right) \right]  \label{t1} \\
&&\mathrm{Tr}_{A}\left[ \hat{f}_{A}\left( s\right) \hat{f}_{A}^{\dag }\left(
t\right) \rho ^{\left( SA\right) }\left( s\right) \right]  \label{t2} \\
&&\mathrm{Tr}_{A}\left[ \hat{f}_{A}^{\dag }\left( t\right) \hat{f}_{A}\left(
s\right) \rho ^{\left( SA\right) }\left( s\right) \right]  \label{t3} \\
&&\mathrm{Tr}_{A}\left[ \hat{f}_{A}^{\dag }\left( s\right) \hat{f}_{A}\left(
t\right) \rho ^{\left( SA\right) }\left( s\right) \right]  \label{t4}
\end{eqnarray}%
with $s\leq t$. Noting that the product of $\hat{f}_{A}\left( t\right) $ ($%
\hat{f}_{A}^{\dag }\left( t\right) $) and $\hat{f}_{A}\left( s\right) $ ($%
\hat{f}_{A}^{\dag }\left( s\right) $) is nonzero only when there exist
integers $n_{t}$ and $n_{s}$ so that $\Lambda _{n_{t}}(t)=\Lambda
_{n_{s}}(s)=1$ or
\begin{eqnarray}
\exists\ n_t,n_s,\ \mathrm{so\ that}\ t\in \left[ t_{n_{t}}^{\prime
},t_{n_{t}}\right] ;s\in \left[ t_{n_{s}}^{\prime },t_{n_{s}}\right].
\label{cond}
\end{eqnarray}
When this condition is satisfied, we separate all the apparatuses into two
groups:
\begin{eqnarray}
&&X:\{\mathrm{the\ apparatuses}\ A_{1},A_{2}...,A_{n_{s}}\}  \notag \\
&&Y:\{\mathrm{the\ apparatuses}\ A_{n_{s}+1},A_{n_{s}+2},...\}  \notag
\end{eqnarray}
Obviously, $X$ is the group of the apparatuses which interact with the
system before the time $s$, while $Y$ includes the ones interact with the
system after the time $s$. Therefore the density matrix $\rho ^{\left(
SA\right) }\left( s\right) $ can be written as
\begin{eqnarray}
\rho ^{\left( SA\right) }\left( s\right) =\rho ^{\left( SX\right) }\left(
s\right) \rho _{0}^{Y},
\end{eqnarray}%
where $\rho ^{\left( SX\right) }\left( s\right) $ is the reduced density
matrix of the TLS together with the apparatuses in the group $X$, and $\rho
_{0}^{Y}$ is given by
\begin{eqnarray}
\rho _{0}^{Y}=\prod_{m=n_{s}+1}^{\infty }|\mathrm{app}\rangle ^{\left(
m\right) }\langle \mathrm{app}|.  \label{rhoy}
\end{eqnarray}%
On the other hand, Eqs.~(\ref{fa}), (\ref{fb}) and (\ref{mn}), together with
the condition $s\leq t$, yield
\begin{eqnarray}
\hat{f}_{A}\left( t\right) \hat{f}_{A}^{\dag }\left( s\right)
=\prod_{l=n_{s}+1}^{n_{t}}\exp \left[ iH_{e}^{\left( l\right) }\tau _{%
\mathrm{M}}\right] \exp \left[ -iH_{g}^{\left( l\right) }\tau _{\mathrm{M}}%
\right] .
\end{eqnarray}%
Namely, the operator $\hat{f}_{A}\left( t\right) \hat{f}_{A}^{\dag }\left(
s\right) $ only operates on the state of the apparatus in the group $Y$.
Then the quantity in Eq.~(\ref{t1}) can be obtained as
\begin{eqnarray}
&&\mathrm{Tr}_{A}\left[ \hat{f}_{A}\left( t\right) \hat{f}_{A}^{\dag }\left(
s\right) \rho ^{\left( SA\right) }\left( s\right) \right]  \notag \\
&=&\rho ^{\left( S\right) }\left( s\right) \mathrm{Tr}_{A}\left[ \hat{f}%
_{A}\left( t\right) \hat{f}_{A}^{\dag }\left( s\right) \rho _{0}^{Y}\right] .
\end{eqnarray}%
{It can} 
be calculated easily with the simple form in Eq.~(\ref{rhoy}) of $\rho
_{0}^{Y}$. The terms in Eqs.~(\ref{t2}-\ref{t4}) can be evaluated in the
similar approach.

Then we get the rate equation {\
\begin{eqnarray}
\frac{d}{dt}P_{e}\left( t\right) =&&-\int_{0}^{t}dsF_{AB}^{\left( +\right)
}\left( t,s\right) P_{e}\left( s\right)  \notag \\
&&+\int_{0}^{t}dsF_{AB}^{\left( -\right) }\left( t,s\right) P_{g}\left(
s\right),  \label{ree}
\end{eqnarray}%
}where $P_{e}\left( t\right) =^{(S)}\!\!\langle e|\rho ^{(S)}(t)|e\rangle
^{\left( S\right) }$ and $P_{g}\left( t\right) =1-P_{e}\left( t\right) $ are
the probabilities of the excited and ground states of the TLS. In the cases of Sec. IV where
the initial state of the TLS is assumed to be $|e\rangle$, the $P_{e}\left( t\right)$
defined here becomes the survival probability defined in (\ref{sp}). In Eq.~(\ref%
{ree}), the effective time-correlation functions $F_{AB}^{\left( \pm \right)
}\left( t,s\right) $ of the environment are defined as
\begin{eqnarray}
F_{AB}^{\left( \pm \right) }\left( t,s\right) =2\mathrm{Re}\left[
g_{B}^{\left( \pm \right) }\left( t-s\right) g_{A}\left( t,s\right) \right],
\label{fabp}
\end{eqnarray}%
where the bare correlation-functions $g_{B}^{\left( \pm \right) }$ of the
heat bath are given by
\begin{eqnarray}
&&g_{B}^{\left( +\right) }\left( t-s\right) =\mathrm{Tr}_{B}\left[ \hat{f}%
_{B}\left( t\right) \hat{f}_{B}^{\dag }\left( s\right) \rho ^{\left(
B\right) }\right] \\
&=&\sum_{\vec{k}}\left\vert g_{\vec{k}}\right\vert ^{2}\left[ \left( n_{\vec{%
k}}+1\right) e^{-i\Delta _{\vec{k}}\left( t-s\right) }+n_{\vec{k}%
}e^{i\left( \omega _{\vec{k}}+\omega _{eg}\right) \left( t-s\right) }\right]
,  \notag \\
&&g_{B}^{\left( -\right) }\left( t-s\right) =\mathrm{Tr}_{B}\left[ \hat{f}%
_{B}^{\dag }\left( t\right) \hat{f}_{B}\left( s\right) \rho ^{\left(
B\right) }\right] \\
&=&\sum_{\vec{k}}\left\vert g_{\vec{k}}\right\vert ^{2}\left[ n_{\vec{k}%
}e^{i\Delta _{\vec{k}}\left( t-s\right) }+\left( n_{\vec{k}}+1\right)
e^{-i\left( \omega _{\vec{k}}+\omega _{eg}\right) \left( t-s\right) }\right]\label{gbm}
\notag
\end{eqnarray}%
with $n_{\vec{k}}=\mathrm{Tr}_{B}[a_{\vec{k}}^{\dag }a_{_{\vec{k}}}\rho
^{\left( B\right) }]$ the average number of the boson in the $\vec{k}$-th
mode of the heat bath. The correlation-function of the measurements is
defined as%
\begin{equation}
g_{A}\left( t,s\right) =\mathrm{Tr}_{A}\left[ \hat{f}_{A}\left( t\right)
\hat{f}_{A}^{\dag }\left( s\right) \rho _{0}^{Y}\right].
\end{equation}

\begin{figure}[tbp]
\includegraphics[bb=51bp 264bp 511bp 596bp,clip,width=7cm]{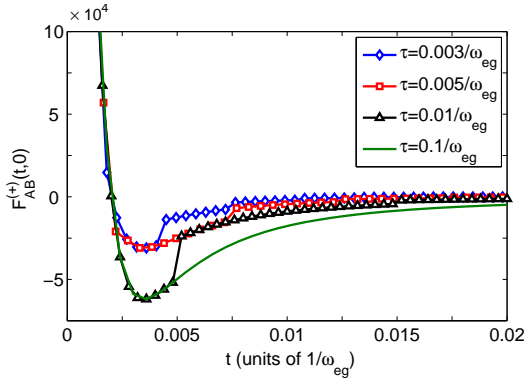}
\caption{(color online) The effective time-correlation function $%
F^{(+)}_{AB}(t,0)$ defined in Eq.~(\ref{fabp}) of the environment of
a dissipative TLS under repeated measurements. Here the heat bath is assumed
to be {at} zero temperature with Ohmic spectrum~(\ref{om}), where $%
\omega_c=500\omega_{eg}$. The incoherence factor of the
non-ideal measurements is $\gamma=0.5$ and $\theta=0$. We
plot the behaviors of $F^{(+)}_{AB}$ with the measurement period $%
\tau=0.003/\omega_{eg}$ (blue line with open diamond), $0.005/%
\omega_{eg}$ (red line with open square), $0.01/\omega%
_{eg}$ (black line with open triangle) and $0.1/\omega_{eg}$
(green line). It is clearly shown that the long-time trail of $%
F^{(+)}_{AB}(t,0)$ is suppressed by the frequent measurements with small $%
\tau$. We also take the limit $\tau_{\mathrm{M}}=0$.}
\end{figure}

It is easy to prove that, the functions $g_{A}\left( t,s\right) $ and $%
g_{B}^{\left( \pm \right) }\left( t-s\right) $ decrease when the absolute
value of $\left\vert t-s\right\vert $ {increases.} 
When $t$ and $s$ satisfy the condition~(\ref{cond}) we have
\begin{eqnarray}
g_{A}\left( t,s\right) =\gamma ^{\left( n_{t}-n_{s}\right) }e^{i\left(
n_{t}-n_{s}\right) \theta }.  \label{gats}
\end{eqnarray}%
If the condition (\ref{cond}) is violated we have $g_{A}\left( t,s\right) =0$%
.

The rate equation~(\ref{ree}) shows that, the evolution of the probabilities
$P_{e,g}\left( s\right) $ of the excited and ground states of the TLS is
governed by the time-correlation functions $F_{AB}^{\left( \pm \right)
}\left( t,s\right) $, which are given by both the time-correlation function
of the heat bath and the decoherence factors of the measurements. The
measurements tune the correlation function $F_{AB}^{\left( \pm \right)
}\left( t,s\right) $ through the function $g_{A}\left( t,s\right) $.
Especially, the trail of $F_{AB}^{\left( \pm \right) }\left( t,s\right) $ in
the long-time-interval region with large $\left\vert t-s\right\vert $ would
be suppressed by the factor $\gamma ^{\left( n_{t}-n_{s}\right) }$ in the
function $g_{A}\left( t,s\right) $ defined in Eq.~(\ref{gats}).

To illustrate the effects given by the repeated measurements to the
effective correlation function $F_{AB}^{\left( \pm \right) }\left(
t,s\right) $, in Fig.~5 we plot $F_{AB}^{\left( +\right) }\left( t,0\right) $
for a TLS in a zero-temperature environment with Ohmic spectrum
\begin{eqnarray}
G\left( \omega \right) =G_{O}\left( \omega \right) =\omega e^{-\omega
/\omega _{c}}.  \label{om}
\end{eqnarray}%
It is clearly shown that when the increasing of the frequency of the
measurements, or the decreasing of the time interval $\tau $ between
measurements, can lead to the suppression of the long-time trail of $%
F_{AB}^{\left( +\right) }\left( t,0\right) $.

One important parameter for the correlation function $F_{AB}^{\left( \pm
\right) }\left( t,s\right) $ is the effective correlation time $\tau _{F}$,
which gives $F_{AB}^{\left( \pm \right) }\left( t,s\right) \simeq 0$ for $%
\left\vert t-s\right\vert >\tau _{F}$. If $\tau _{F}$ is small enough so
that the variation of the probabilities $P_{e}\left( t\right) $ and $%
P_{g}\left( t\right) $ is negligible in the time interval $\tau _{F}$, we
can perform the time-local approximation
\begin{eqnarray}
P_{e,g}\left( s\right) \approx P_{e,g}\left( t\right)  \label{tla}
\end{eqnarray}%
in Eq.~(\ref{ree}) and significantly simplify the rate equation.

\begin{figure}[tbp]
\includegraphics[bb=38bp 261bp 517bp 591bp,clip,width=7cm]{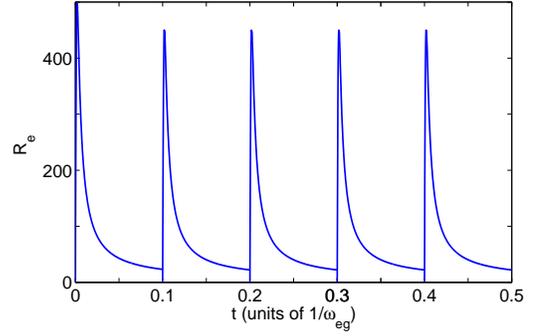}
\caption{(color online) The decay rate $R_{e}(t)$ in Eq.~(\ref{ret}%
). Here we take the measurement period $\tau=0.1/\omega_{eg}$%
. Other parameters are the same as the ones in Fig.~5. As proved in Eq.~(%
\ref{period}), in the long-time limit the decay rate is a periodic
function of $\tau$. (Here we also take the limit $\tau_{%
\mathrm{M}}=0$.) The discontinuity of the function $R_{e}(t)$ at $t=n%
\tau\ (n=1,2,3,...)$ is due to the jumping behavior of the function $%
g_{A}\left( t,s\right)$ in Eq.~(\ref{gats}).}
\end{figure}

Eq.~(\ref{fabp}) yields
\begin{eqnarray}
\tau _{F}=\min \left\{ \tau _{A},\tau _{B}\right\},  \label{tauf}
\end{eqnarray}%
where $\tau _{A}$ and $\tau _{B}$ are the correlation times of the functions
$g_{A}\left( x\right) $ and $g_{B}^{\left( \pm \right) }$. From Eq. (\ref%
{gats}) we have
\begin{eqnarray}
\tau _{A}\sim \left( 1-1/\ln \gamma \right) \left( \tau +\tau _{\mathrm{M}%
}\right) .
\end{eqnarray}%
In the case of ideal projective measurements we have $\tau _{A}=\tau +\tau _{%
\mathrm{M}}$. Therefore, Eq.~(\ref{tauf}) shows that frequent measurements
with small period $\tau$ can reduce the effective correlation-time of the
environment experienced by the dissipative TLS.

\subsection{The master equation under time-local and coarse-grained
approximation}

In the following we assume the effective correlation time $\tau _{F}$ is
small enough and the time-local approximation~(\ref{tla}) can be used. Then
the rate equation~(\ref{ree}) can be simplified as
\begin{eqnarray}
\frac{d}{dt}P_{e}\left( t\right) =-R_{e}\left( t\right) P_{e}\left( t\right)
+R_{g}\left( t\right) P_{g}\left( t\right),
\end{eqnarray}%
where the time-dependent decay rates $R_{e}\left( t\right) $ and $%
R_{g}\left( t\right) $ are given by
\begin{eqnarray}
R_{e,g}\left( t\right) =\int_{0}^{t}dsF_{AB}^{\left( \pm \right) }\left(
t,s\right).  \label{ret}
\end{eqnarray}

On the other hand, Eq.~(\ref{gats}) implies that
\begin{eqnarray}
g_{A}\left( t,s\right) =g_{A}\left( t+\tau +\tau _{\mathrm{M}},s+\tau +\tau
_{\mathrm{M}}\right) .
\end{eqnarray}%
This relationship, together with the definitions of $F_{AB}^{\left( \pm
\right) }\left( t,s\right) $, gives%
\begin{eqnarray}
F_{AB}^{\left( \pm \right) }\left( t,s\right) =F_{AB}^{\left( \pm \right)
}\left( t+\tau +\tau _{\mathrm{M}},s+\tau +\tau _{\mathrm{M}}\right) .
\end{eqnarray}%
Therefore, when $t$ is much longer than the effective correlation time $\tau
_{F}$, we have%
\begin{eqnarray}
R_{e,g}\left( t\right) &=&\int_{0}^{t}dsF_{AB}^{\left( \pm \right) }\left(
t,s\right)  \notag \\
&=&\int_{0}^{t}dsF_{AB}^{\left( \pm \right) }\left( t+\tau +\tau _{\mathrm{M}%
},s+\tau +\tau _{\mathrm{M}}\right)  \notag \\
&=&\int_{\tau +\tau _{\mathrm{M}}}^{t+\tau +\tau _{\mathrm{M}%
}}dsF_{AB}^{\left( \pm \right) }\left( t+\tau +\tau _{\mathrm{M}},s\right)
\notag \\
&\approx &\int_{0}^{t+\tau +\tau _{\mathrm{M}}}dsF_{AB}^{\left( \pm \right)
}\left( t+\tau +\tau _{\mathrm{M}},s\right)  \notag \\
&=&R_{e,g}\left( t+\tau +\tau _{\mathrm{M}}\right),  \label{period}
\end{eqnarray}%
and then the decay rates become periodic functions of $t$ with period $\tau
+\tau _{\mathrm{M}}$, which is the same as the period of the measurements.
In {Fig.~6} 
we plot the $R_{e}\left( t\right) $ for the system in the calculation of {%
Fig.~5.} 
The periodic behavior of $R_{e,g}\left( t\right)$ in the large $t$ case is
illustrated clearly.

If the measurements are frequent enough so that the variation of the
probabilities $P_{e,g}\left( t\right) $ in the time interval $\tau +\tau _{%
\mathrm{M}}$ can be neglected, we can further perform the coarse-grained
approximation and obtain the Markovian rate equation
\begin{eqnarray}
\frac{d}{dt}P_{e}\left( t\right) =-R_{e}^{\mathrm{CG}}P_{e}\left( t\right)
+R_{g}^{\mathrm{CG}}P_{g}\left( t\right) \label{rateequation}
\end{eqnarray}%
with the coarse-grained decay rates
\begin{eqnarray}
R_{e,g}^{\mathrm{CG}}=\lim_{N\rightarrow \infty }\frac{1}{N\left( \tau +\tau
_{\mathrm{M}}\right) }\int_{0}^{N\left( \tau +\tau _{\mathrm{M}}\right)
}R_{e,g}\left( t\right) dt.
\end{eqnarray}%
It is easy to prove that in the zero-temperature case we have
\begin{eqnarray}
R_{e}^{\mathrm{CG}}&=&\lim_{t_F\rightarrow \infty }\frac{1}{t_F}%
\int_0^{t_F}dt\int_0^tds g_B^{(+)}(t-s)g_A(t,s)+h.c.  \notag \\
&=&\lim_{t_F\rightarrow \infty }\frac{1}{t_F}\int_0^{t_F}dt\int_0^{t_F}ds\left[%
g_B^{(+)}(t-s)g_A(t,s)\right]  \notag \\
&=&\lim_{t_F\rightarrow \infty }\frac{1}{t_F}\left|\int_0^{t_F}dt
H_I(t)|\Psi(0)\rangle_I\right|^2
\end{eqnarray}
with $H_I(t)$ and $|\Psi(0)\rangle_I$ defined in Eqs.~(\ref{hit}) and (\ref%
{psi0}). Therefore $R_{e}^{\mathrm{CG}}$ at zero temperature is reduced to the short-time
decay rate $R_{\mathrm{mea}}(\tau )$ being defined in Eq.~(\ref{rmea2}). Namely, the
decay rate given by calculation for {the short-time} 
evolution also governs the long-time evolution when the effective
correlation time $\tau _{F}$ and the period of the measurements $\tau +\tau
_{\mathrm{M}}$ is short enough.

Finally we discuss the behavior of the coarse-grained decay rate $R_{g}^{\rm CG}$
of the ground state.
It is well-known that, for a dissipative TLS without measurements, the decay rate of the ground state
is usually negligible in the zero-temperature case. However, due to the counter-rotating terms
of the Hamiltonian (\ref{hf}), the correlation-functions $g_{B}^{\left( \pm \right) }$
defined in (\ref{gbm}) and the ground-state decay rate are not absolutely zero.
When the periodical measurements are performed,
the value of the ground-state decay rate $R_{g}^{\rm CG}$ is also varied by the measurements, and
can take significant value even in the zero-temperature.

For simplicity, we consider the simple cases
with periodic identical measurements
which has the decoherence factor $\gamma e^{i \theta}\ (\gamma<1)$, i.e.,
the cases discussed in Sec. IV A. The straightforward calculation shows that in
such a case the coarse-grained
decay rates are given by
\begin{eqnarray}
R_{e,g}^{\rm CG}\left( \tau \right) =\int d\eta G\left( \frac{\eta }{\tau }%
\pm\omega _{eg}\right) h\left(\gamma ,\theta -\eta \right)\mathrm{sinc}^{2}
\frac{\eta }{2},  \label{rmea3}
\end{eqnarray}
where ``$+$" stands for $R_{e}^{\rm CG}$ and ``$-$" for $R_{g}^{\rm CG}$. In
the above expression, $\tau$ is length of the time interval between two measurements.
The function $h$ and the spectrum $G$ of the heat bath are defined in Eq. (\ref{h}) and
(\ref{g}).

Therefore, when there is no measurements, or in the limit $\tau \rightarrow \infty$, we have
the decay rates of ground state and excited state
\begin{eqnarray}
R_{e,g}^{\rm CG}(\tau\rightarrow \infty)\approx G\left( %
\pm\omega _{eg}\right)\int d\eta  h\left(\gamma ,\theta -\eta \right)\mathrm{sinc}^{2}
\frac{\eta }{2}.
\end{eqnarray}
Since all the frequencies of the heat-bath are positive, we have $G\left( %
-\omega _{eg}\right)=0$ and then $R_{g}^{\rm CG}(\tau\rightarrow \infty)\approx 0$.

\begin{figure}[tbp]
\includegraphics[bb=38bp 198bp 506bp 644bp,clip,width=8.5cm]{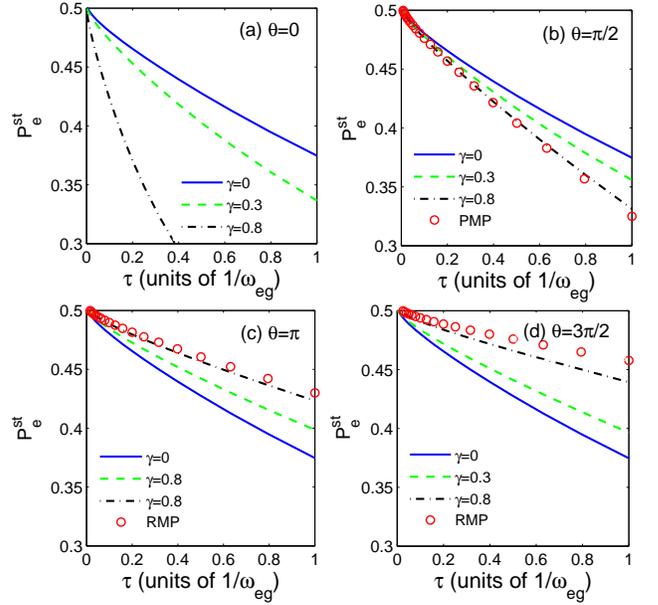}
\caption{ (color online) The probability
$P_{e}^{\rm st}$ of the excited state of the TLS in the steady-state.
Here we plot $P_{e}^{\rm st}$ given by Eq.~(\ref{pest}). We take the noise spectrum in Eq.~(\ref{gh}) and
illustrate the cases with (a) $\theta=0$, (b) $\theta=\pi%
/2$, (c) $\theta=\pi$ and (d) $\theta=3\pi/2$, and $\gamma=0$ (blue solid line), $\gamma=0.3$ (green
dashed line), $\gamma=0.8$ (black dashed-dotted line). In (b-d) we also plot the %
probability
$P_{e}^{\rm st}$ (red open circle) given by frequent phase modulation pulses.}
\end{figure}

In the presence of periodic measurements, the decay rates $R_{e,g}^{\rm CG}$ are given by the overlap
of the spectrum $G\left(\eta/\tau \pm\omega _{eg}\right)$ and the function $h\left(\gamma ,\theta -\eta \right)\mathrm{sinc}^{2}
(\eta/2)$, or, roughly speaking, given by the values of
$G\left(\eta/\tau \pm\omega _{eg}\right)$ in the region $\eta\in[-\pi,\pi]$. We assume
the function $G(x)$ takes nonzero value in the region $x\in[0,\Omega]$, then
$G\left(\eta/\tau \pm\omega _{eg}\right)$ is nonzero only when $\eta\in[\mp\omega_{eg}\tau,\mp\omega_{eg}\tau+\Omega\tau]$.
Therefore, when the measurements are more frequent,
 or the time interval $\tau$ becomes smaller, the non-zero region of
 $G\left(\eta/\tau -\omega _{eg}\right)$ has more and more overlaps with the region
 $[-\pi,\pi]$. Then the decay rates $R_{g}^{\rm CG}$ can be significant.
 In the limit $\tau\rightarrow 0$, both of the two functions $G\left(\eta/\tau \pm\omega _{eg}\right)$
take non-zero values only in a small region around $\eta=0$. Then we have
\begin{eqnarray}
R_{g}^{\rm CG}(\tau\rightarrow 0)=R_{g}^{\rm CG}(\tau\rightarrow 0)\approx \tau h(\gamma,\theta)
\int d\xi G\left(\xi\right).
\end{eqnarray}

The influence of the finite value of $R_{g}^{\rm CG}$ can be observed from the steady-state solution
of the coarse-grained rate equation (\ref{rateequation}), which describes the result of the long-time
evolution of the dissipative TLS under repeated measurements. According to Eq. (\ref{rateequation}),
in the steady state, population probabilities
$P_{g}^{\rm st}$ and $P_{e}^{\rm st}$ of the TLS in the ground and excited states
can be expressed as a function of the time interval $\tau$ of the measurement:
\begin{eqnarray}
P_{g}^{\rm st}(\tau)=\frac{R_{e}^{\rm CG}(\tau)}{R_{e}^{\rm CG}(\tau)+R_{g}^{\rm CG}(\tau)};
\ P_{e}^{\rm st}=\frac{R_{g}^{\rm CG}(\tau)}{R_{e}^{\rm CG}(\tau)+R_{g}^{\rm CG}(\tau)}.\nonumber\\
\label{pest}
\end{eqnarray}
Therefore, the non-zero decay rate $R_{g}^{\rm CG}$ of the ground state
 leads to the non-zero population probability
$P_{e}^{\rm st}$ of the excited state.

To illustrate the effects given by finite $R_{g}^{\rm CG}$,
in Fig. 7(a)-7(d) we plot the population probabilities $P_{e}^{\rm st}$
with respect to different values of $\tau$ and the decoherence factor $\gamma e^{i\theta}$.
As in Fig. 4, we take the noise spectrum in Eq.~(\ref{gh}).
It is clearly shown that when $\tau$ becomes small, the probability $P_{e}^{\rm st}$
becomes non-zero. In the limit $\tau\rightarrow 0$, $P_{e}^{\rm st}$ approaches $1/2$
which implies $R_{g}^{\rm CG}=R_{e}^{\rm CG}$. In Fig. 7
we also plot $P_{e}^{\rm st}$ for the cases with phase modulation pulses rather than
repeated measurements. The behavior of $P_{e}^{\rm st}$ is
quit similar with the one given by repeated measurements. The non-zero
decay rate $R_{g}^{\rm CG}$ for the cases with phase modulation pulses
is also obtained in Refs \cite{phaseM2,phaseM3}.

\section{Conclusion and Discussion}

In this paper we provide a complete dynamical model for the evolution of
dissipative TLS under repeated quantum non-demolition measurements. Our model gives an explanation
of QZE and QAZE without the wave function collapse postulation.
The effects given by non-ideal measurements are naturally obtained
in our model. The QZE and QAZE given by repeated phase modulation pulses
can also be derived in our framework as a special case of non-ideal measurement.

Based on our model, we derive the short-time decay rate~(\ref{bigr}) which
implies that the QZE and QAZE may be enhanced by a non-ideal measurement
with complex decoherence factor. The long-time rate equation~(\ref{ree}) is
also obtained in terms of the effective time-correlation function $%
F_{AB}^{(\pm)}$, which describes the adjustment of the noise spectrum from
the repeated measurements. The rate equation also shows that, the decay rate
of the ground state of the TLS may be changed to non-zero value by the repeated measurements,
and then the steady-state probabilities
of the ground state and the excited state
are also varied.

The effects of non-ideal measurements
on QZE and QAZE obtained in our model can be observed in the experiments where the
system-apparatus interactions are well controlled. Such systems
are possibly to be realized by nuclear magnetic resonance
or solid-state quantum devices. Although we illustrate our
model with a TLS, all the techniques in this paper, including the choice of
the interaction picture in Sec. II.C, the time-dependent perturbation theory
in Sec. III, the master equation~(\ref{me1}) and the separation of the
apparatuses in Sec. IV.A, can be used in the cases other than TLS. Therefore
our model presented in this paper can also be straightforwardly generalized
to the discussions of QZE and QAZE of multi-level quantum system.

\begin{acknowledgments}
This work was supported by National Natural Science Foundation of China {%
under} 
Grant Nos. 11074305, 10935010, 11074261 and the Research Funds of Renmin
University of China (10XNL016). One of the authors (Peng Zhang) appreciates
Dr. Ran Zhao for helpful discussions.
\end{acknowledgments}

\end{document}